\documentclass[prl,twocolumn,showpacs,preprintnumbers,amsmath,amssymb]{revtex4}

\usepackage{graphicx}
\usepackage{dcolumn}
\usepackage{bm}

\newcommand{\mapright}[1]{\smash{\mathop{\hbox to 1.0cm{\rightarrowfill}}\limits^{#1}}}

\begin{document}


\title{Theory of Macroscopic Quantum Tunneling and Dissipation in High-$T_c$ Josephson Junctions}

\author{Shiro Kawabata,$^{1,2}$ Satoshi Kashiwaya,$^3$ Yasuhiro Asano,$^4$ Yukio Tanaka,$^5$ Takeo Kato,$^6$and Alexander A. Golubov$^1$}
\affiliation{%
$^1$Faculty of Science and Technology, University of Twente, 
P.O. Box 217, 7500 AE Enschede, The Netherlands \\
$^2$Nanotechnology Research Institute (NRI), National Institute of 
Advanced Industrial Science and Technology (AIST), Tsukuba, 
Ibaraki, 305-8568, Japan \\
$^3$Nanoelectronics Research Institute (NeRI), National Institute of 
Advanced Industrial Science and Technology (AIST), Tsukuba, 
Ibaraki, 305-8568, Japan \\
$^4$Department of Applied Physics, Hokkaido University,
Sapporo, 060-8628, Japan\\
$^5$Department of Applied Physics, Nagoya University,
Nagoya, 464-8603, Japan\\
$^6$The Institute for Solid State Physics (ISSP), University of Tokyo,
Kashiwa, 277-8581, Japan
}%

\date{\today}

\begin{abstract}
We have investigated macroscopic quantum tunneling (MQT) in in-plane high-$T_c$ superconductor Josephson junctions and the influence of the nodal-quasiparticle and the zero energy bound states (ZES) on MQT.
We have shown that the presence of the ZES at the interface between the insulator and the superconductor leads to strong Ohmic quasiparticle dissipation.
Therefore, the MQT rate is noticeably suppressed in comparison with the $c$-axis junctions in which ZES are completely absent.
\end{abstract}

\pacs{}
\maketitle

\section{Introduction}

A mesoscopic single Josephson junction is an interesting physical object for testing quantum mechanics at a macroscopic level. 
In current-biased Josephson junctions, the measurements of macroscopic quantum tunneling (MQT) are performed by switching the junction from its metastable zero-voltage state to a non-zero voltage state (see Fig.~1 (d)).
Until now, experimental investigations of MQT have been focused on $s$-wave (low-$T_c$) junctions only.
This fact is due to the naive preconception that the existence of the low energy quasiparticles in the $d$-wave order parameter of a high-$T_c$ cuprate superconductor~\cite{rf:d-wave2} may preclude the possibility of observing the MQT.

Recently we have theoretically investigated the effect of the nodal-quasiparticle on MQT in the $d$-wave $c$-axis junctions (e.g., Bi2212 intrinsic Josephson junctions~\cite{rf:Intrinsic,rf:You} and cross whisker junctions~\cite{rf:Takano})~\cite{rf:Kawabata1,rf:Kawabata1-2}.
We have shown that the effect of the nodal-quasiparticle gives rise to a super-Ohmic dissipation~\cite{rf:d-waveaction1,rf:d-waveaction2} and the suppression of the MQT due to the nodal-quasiparticle is very weak. 

The first experimental observation of the MQT in the high-$T_c$ Josephson junction was made by Bauch $et$ $al$., using a YBCO grain boundary bi-epitaxial junction~\cite{rf:Bauch1,rf:Bauch2}.
Recently, Inomata $et$ $al.$~\cite{rf:Inomata},
 Jin $et$ $al.$~\cite{rf:Jin},
 and Kashiwaya $et$ $al.$~\cite{rf:Kashiwaya1,rf:Kashiwaya2}
 have experimentally observed the MQT in the $c$-axis (Bi2212 intrinsic) junctions.
They reported that the effect of the nodal-quasiparticle on the MQT is negligibly small and the thermal-to-quantum crossover temperature is relatively high (0.5$\sim$1K) compared with the case of the low-$T_c$ and the YBCO bi-epitaxial junctions.
In Jin $et$ $al.$' s experiment, $O(N^2)$ ($N$ is the number of the stacked junctions)  enhancement of the MQT rate was reported.
This enhancement is attributed to collective motion of the phase differences in the intrinsic junctions~\cite{rf:Machida,rf:Fistul,rf:Nori}.

%
%
%
%
\begin{figure}[b]
\begin{center}
\includegraphics[width=8.0cm]{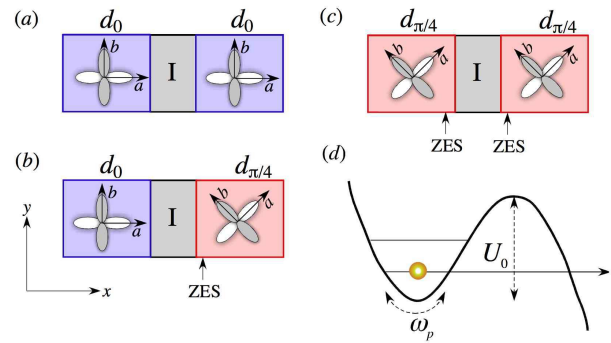}
\end{center}
\caption{Schematics of the in-plane $d$-wave Josephson junction. (a) $d_0/d_0$, (b) $d_0/d_{\pi/4}$, and (c) $d_{\pi/4}/d_{\pi/4}$ junction. 
In the case of the $d_0/d_{\pi/4}$ and $d_{\pi/4}/d_{\pi/4}$ junctions, the ZES are formed near the boundary between superconductor $d_{\pi/4}$ and insulating barrier I. 
(d) Potential $U(\phi)$ v.s. the phase difference  $\phi$ between two superconductors.
$U_0$ is the barrier height and $\omega_p$ is the Josephson plasma frequency.
}
\label{f1}
\end{figure}
%
%

In this paper, we will theoretically investigate the MQT in the $d$-wave in-plane junctions parallel to the $ab$-plane (see Fig.~1)~\cite{rf:Kawabata2}.
In such junctions, the zero energy bound states (ZES)~\cite{rf:KashiwayaTanaka} are formed near the interface between superconductor and the insulating barrier.
ZES are generated by the combined effect of multiple Andreev reflections and the sign change of the $d$-wave order parameter symmetry, and are bound states for the quasiparticle at the Fermi energy.
Below, we will show that ZES give rise to Ohmic type strong dissipation so MQT is considerably suppressed in compared with the $c$-axis and the $d_0/d_0$ junction cases.

\section{Effective Action}
By using the method developed by Eckern $et$. $al$.,~\cite{rf:Schon1} the partition function of the system can be described by a functional integral over the macroscopic variable (the phase difference $\phi$), 
\begin{eqnarray}
 Z 
= 
\int 
 {\cal D} \phi (\tau) 
\exp
\left(
  - \frac{S_{eff}[\phi]}{\hbar}
\right)
.
\end{eqnarray}
In the high barrier limit, $i. e.,$ $z_0\equiv m w_0/\hbar^2 k_F  \gg 1$( $m$ is the mass of the electron, $w_0$ is the height of the delta function potential I, and $k_F$ is the Fermi wave length), the effective action $S_{eff}$ is given by 
\begin{eqnarray}
S_{eff}[\phi]
&= &
\int_{0}^{\hbar \beta} d \tau 
\left[
   \frac{M}{2} 
   \left(
   \frac{\partial \phi ( \tau) }{\partial \tau}
   \right)^2
   + 
   U(\phi)
\right] + S_\alpha[\Phi],
\nonumber\\
S_\alpha[\Phi]
&=&
-
\int_{0}^{\hbar \beta} d \tau 
 \int_{0}^{\hbar \beta} d \tau'
  \alpha (\tau - \tau') \cos \frac{\phi(\tau) - \phi (\tau') }{2}
.
 \nonumber\\
\end{eqnarray}
In this equation, $\beta = 1 /k_B T$, $M
 = 
 C \left( \hbar/2 e\right)^2
$
is the mass ($C$ is the capacitance of the junction) and the potential  $U(\phi)$ is described by
\begin{eqnarray}
 U(\phi) 
 = 
  \frac{\hbar}{2 e} 
\left[
    \int_0^1 d \lambda \ \phi  I_J (\lambda \phi) - \phi \  I_{ext}
\right]
,
\end{eqnarray}
where $I_J$ is the Josephson current and $ I_{ext}$ is the external bias current, respectively.
The dissipation kernel $\alpha(\tau)$ is related to the quasiparticle current $I_{qp}$
under constant bias voltage $V$ by
\begin{eqnarray}
\alpha(\tau) 
=
\frac{\hbar  }{e}
\int_0^\infty\frac{d \omega}{2 \pi} e^{-\omega \tau} 
    I_{qp} \left( V=\frac{\hbar \omega}{e }\right) 
,
\end{eqnarray}
at zero temperature.

Below, we will derive the effective action for the three types of the $d$-wave junction ($d_0/d_0$, $d_0/d_{\pi/4}$, and $d_{\pi/4}/d_{\pi/4}$) in order to investigate the effect of the nodal-quasiparticles and ZES on MQT.
In the case of the $d_0/d_0$ junction, the node-to-node quasiparticle tunneling can contribute to the dissipative quasiparticle current.
However, ZES are completely absent.
These behaviors are qualitatively identical with that for the $c$-axis Josephson junctions~\cite{rf:Kawabata1,rf:Kawabata1-2}.
On the other hand, in the case of the $d_0/d_{\pi/4}$ and $d_{\pi/4}/d_{\pi/4}$ junction, the ZES are formed around the surface of the superconductor $d_{\pi/4}$.
Therefore the node to ZES ($d_0/d_{\pi/4}$) and the ZES to ZES ($d_{\pi/4}/d_{\pi/4}$) quasiparticle tunneling becomes possible.

 Firstly, we will calculate the potential energy $U$ in the effective action~(2).
 As mentioned above, $U$ can be described by the Josephson current through the junction in the high barrier limit.
 In order to obtain the Josephson current we start from the  Bogoliubov-de Gennes (B-dG) equation~\cite{rf:KashiwayaTanaka},

\begin{eqnarray}
&&
\! \! \! \! \! 
\int d \mbox{\boldmath $r$}'
\left(
\begin{array}{cc}
\delta(\mbox{\boldmath $r$} -\mbox{\boldmath $r$}' ) h(\mbox{\boldmath $r$}' )
& 
\Delta(\mbox{\boldmath $r$} -\mbox{\boldmath $r$}' )e^{i \varphi}
\\
\Delta^*(\mbox{\boldmath $r$} -\mbox{\boldmath $r$}' )e^{-i \varphi}
 &  
-\delta(\mbox{\boldmath $r$} -\mbox{\boldmath $r$}' ) h^*( \mbox{\boldmath $r$}' )
\end{array}
\right)
\left(
\begin{array}{c}
u({ \mbox{\boldmath $r$}}) \\
v({ \mbox{\boldmath $r$}})
\end{array}
\right)
\nonumber\\
&=&E
\left(
\begin{array}{c}
u({ \mbox{\boldmath $r$}}) \\
v({ \mbox{\boldmath $r$}})
\end{array}
\right)
,
 \end{eqnarray}
where $\varphi$ is the phase of order parameter, $u(v)$ is the amplitude of the wave function for the electron (hole)-like quasiparticle, 
$
 h(\mbox{\boldmath $r$})
= - \hbar^2\nabla^2/2m -\mu+w_0 \delta(x)
$, and
$
\Delta(\mbox{\boldmath $r$} -\mbox{\boldmath $r$}' ) 
=\Omega^{-1} \sum_{\mbox{\boldmath $k$}} \Delta_{\mbox{\boldmath $k$}}
\exp \left[  i \mbox{\boldmath $k$} \cdot (\mbox{\boldmath $r$} -\mbox{\boldmath $r$}' ) \right]
$
is the order parameter ($\Omega$ is the volume of the superconductor).
In the superconductor regions ($d_0$ and $d_{\pi/4}$), the B-dG equation (5) can be transformed into the eigenequation
\begin{eqnarray}
\left(
\begin{array}{cc}
\xi_{ \mbox{\boldmath $k$}} & \Delta_{ \mbox{\boldmath $k$}} e^{i \varphi}\\
 \Delta_{ \mbox{\boldmath $k$}} e^{-i \varphi}&  -\xi_{ \mbox{\boldmath $k$}}
\end{array}
\right)
\left(
\begin{array}{c}
u_{ \mbox{\boldmath $k$}} \\
v_{ \mbox{\boldmath $k$}}
\end{array}
\right)
=E
\left(
\begin{array}{c}
u_{ \mbox{\boldmath $k$}} \\
v_{ \mbox{\boldmath $k$}}
\end{array}
\right)
,
 \end{eqnarray}
where, $\xi_{ \mbox{\boldmath $k$}}=\hbar^2 k^2/2m + \hbar^2 p^2/2m  - \mu \ $($p=2 \pi n/D$ and $D$ is the width of the junction).
The amplitude of the order parameter $\Delta_{\mbox{\boldmath $k$}}$ is given by $\Delta_0 \cos 2 \theta \equiv \Delta_{d_0}(\theta)$ for $d_0$ and $\Delta_0 \sin 2 \theta \equiv \Delta_{d_{\pi/4}}(\theta)$ for $d_{\pi/4}$, where $\cos \theta = k / k_F$.
The Andreev reflection coefficient for the electron (hole)-like quasiparticle $r_{he}$ ($r_{eh}$) is calculated by solving the eigenequation (6) together with the appropriate boundary conditions.
By substituting $r_{he}(r_{eh})$ into the formula of the Josephson current for unconventional superconductors (the Tanaka-Kashiwaya formula)~\cite{rf:KashiwayaTanaka},
\begin{eqnarray}
I_J=
\frac{e  }{\hbar}
\sum_{p}
\frac{1}{\beta} \sum_{\omega_n}
\left(
\frac{\Delta_+}{\Omega_+} r_{he}
-
\frac{\Delta_-}{\Omega_-} r_{eh}
\right)
,
 \end{eqnarray}
we can obtain $\phi$ dependence of the Josephson current.
Here, $\Delta_\pm=\Delta_{(\pm k,p)}$, $\Omega_\pm = \sqrt{(\hbar \omega_n)^2 - |\Delta_\pm|^2}$, $\omega_n=(2n+1)\pi/\beta \hbar$ is the fermionic Matsubara frequency.
In the case of low temperatures ($\beta^{-1}\ll \Delta_0$) and the high barrier limit ($z_0 \gg 1$), we get 
\begin{eqnarray}
 I_J(\phi) 
\approx 
\left\{
\begin{array}{cl}
\displaystyle{ I_1 \sin \phi}
 \quad 
 & \mbox{for} \quad  \mbox{$d_{0}/d_0$}
 \\
\displaystyle{- I_2 \sin2 \phi }
\quad
&
\mbox{for} \quad   \mbox{$d_{0}/d_{\pi/4}$}
\\
\displaystyle{I_3 \sin \frac{ \phi}{2} }
\quad
&
\mbox{for} \quad   \mbox{$d_{\pi/4}/d_{\pi/4}$}
\\\end{array}
\right.
,
\end{eqnarray}
where $I_1 \equiv 3 \pi \Delta_0 / 10 e R_N$, $I_2 \equiv \pi^2 \hbar \beta \Delta_0^2 / 35 e^3 N_c R_N^2$, and $I_3 \equiv 3 \pi z_0 \Delta_0 /4 e R_N$ ($R_N=3 \pi \hbar z_0^2/ 2 e^2 N_c$ is the normal state resistance of the junction and $N_c $ is the number of channel at the Fermi energy).

By substituting the Josephson current into eq.~(3), we can obtain  the analytical expression of the potential $U,$ i.e.,
\begin{eqnarray}
U(\phi) 
\approx 
\left\{
\begin{array}{cl}
\displaystyle{- \frac{\hbar I_1}{2e}\left(  \cos \phi + \frac{I_{ext}}{I_1}  \phi \right) }
&    
\mbox{for} \ \   \mbox{$d_{0}/d_0$}
 \\
\displaystyle{- \frac{\hbar I_2}{4e}\left( -  \cos  2 \phi + 2 \frac{I_{ext}}{I_2} \phi  \right)}    
&    
\mbox{for} \ \   \mbox{$d_{0}/d_{\pi/4}$}
 \\
\displaystyle{- \frac{\hbar I_3}{e}\left(  \cos   \frac{\phi}{2}  + \frac{1}{2} \frac{I_{ext}}{I_3} \phi  \right)}    
&    
\mbox{for} \ \   \mbox{$d_{\pi/4}/d_{\pi/4}$}
\end{array}
\right.
.
\end{eqnarray}
As in the case of the $s$-wave and the $c$-axis junctions~\cite{rf:Kawabata1}, $U$ can be expressed as the tilted washboard potential (see Fig.~1(d)).

\section{Dissipation due to nodal-quasiparticles and ZES}

Next we will calculate the dissipation kernel $\alpha(\tau)$ in the effective action (2).
In the high barrier limit, the quasiparticle current is given by~\cite{rf:KashiwayaTanaka}
\begin{eqnarray}
I_{qp}(V)
\! 
 &=& 
 \! 
 \frac{2e}{h} \sum_{p} |t_N|^2 \int_{-\infty}^{\infty}dE N_{L} (E,\theta)  N_{R} (E+ eV, \theta) 
 \nonumber\\
 &\times&
\left[
 f(E) - f(E +eV)
\right]
,
\end{eqnarray}
where $t_N \approx \cos \theta /z_0$ is the transmission coefficient of the barrier I, $N_{L(R)} (E,\theta)$ is the quasiparticle surface density of states ($L=d_0$ and $R=d_0$ or $d_{\pi/4}$) and $f(E)$ is the Fermi-Dirac distribution function.
The explicit expression of the surface density of states was obtained by Matsumoto and Shiba~\cite{rf:Matsumoto}.
In the case of $d_0$, there are no ZES.
Therefore the angle $\theta$ dependence of $N_{d_0}(E,\theta)$ is the same as the bulk and is given by
\begin{eqnarray}
 N_{d_{0}} (E,\theta)
=     \mbox{Re} \frac{|E|}{\sqrt{E^2-\Delta_{d_0}(\theta)^2}}
.
\end{eqnarray}
On the other hand, $N_{d_{\pi/4}}(E,\theta)$ is given by 
\begin{eqnarray}
\!
 N_{d_{\pi/4}} (E,\theta)
\!
=
\!
     \mbox{Re} \frac{\sqrt{E^2 \! - \! \Delta_{d_{\pi/4}}(\theta)^2}}{|E|} 
     \!
     +
     \!
      \pi |\Delta_{d_{\pi/4}}(\theta)| \delta(E)
.
\end{eqnarray}
The delta function peak at $E=0$ corresponds to the ZES. 
Because of the bound state at $E=0$, the quasiparticle current for the $d_0/d_{\pi/4}$ and $d_{\pi/4}/d_{\pi/4}$ junctions is drastically different from that for the $d_0/d_0$ junctions in which no ZES are formed.
By substituting Eqs.~(11) and (12) into Eq.~(10), we can obtain the analytical expression of the quasiparticle current $I_{qp}(V)$.
In the limit of low bias voltages ($e V \ll \Delta_0$) and low temperatures ($\beta^{-1} \ll \Delta_0$),  this can be approximated as 
\begin{eqnarray}
I_{qp}(V) 
\approx 
\left\{
\begin{array}{cl}
\displaystyle{\frac{3^2 \pi^2}{2^8 \sqrt{2}} \frac{e V^2}{ \Delta_0 R_N } }
& 
\quad    \mbox{for} \quad \mbox{$d_{0}/d_{0}$}\\
\displaystyle{\frac{3 \pi^2}{2^4 \sqrt{2} } \frac{V}{  R_N } 
   }
& 
\quad   \mbox{for}\quad   \mbox{$d_{0}/d_{\pi/4}$}\\
\displaystyle{\frac{2^5 \pi}{35 } \left(\frac{\Delta_0}{\epsilon}\right)^2  \frac{V}{  R_N } 
   }
& 
\quad   \mbox{for}\quad   \mbox{$d_{\pi/4}/d_{\pi/4}$}
\end{array}
\right.
.
\end{eqnarray}
In the calculation of $I_{qp}$ for the  $d_{\pi/4}/d_{\pi/4}$ junctions, we have replaced the ZES delta function $\delta(E)$ in Eq.~(12) with the Lorentz type function, i.e.,  
\begin{eqnarray}
\delta(E) \to \frac{1}{\pi}\frac{\epsilon}{\epsilon^2 + E^2 },
\end{eqnarray}
in order to avoid a mathematical difficulty and model the real systems (which include e.g. disorder and many body effects).
It is apparent from Eq.~(13) that, in the case of $d_0/d_0$ junctions, the dissipation is of the super-Ohmic type as in the case of the $c$-axis junctions~\cite{rf:Kawabata1}.
This can be attributed to the effect of the node-to-node quasiparticle tunneling.
Thus the quasiparticle dissipation is very weak.
On the other hand, in the case of the $d_0/d_{\pi/4}$ junctions, the node-to-ZES  quasiparticle tunneling gives the Ohmic dissipation which is similar to that in normal junctions~\cite{rf:Schon1}.
Therefore the dissipation for the $d_0/d_{\pi/4}$ junctions is stronger than that for the $d_0/d_{0}$ junctions.
Moreover, in the case of the $d_{\pi/4}/d_{\pi/4}$ junctions, the ZES-to-ZES quasiparticle tunneling dominates the quasiparticle dissipation.
The broadening of the ZES peak $\epsilon$ is typically one order magnitude smaller than 
$\Delta_0$.
Therefore, due to the prefactor $(\Delta_0/\epsilon)^2$ in Eq.~(12), the quasiparticle dissipation in the $d_{\pi/4}/d_{\pi/4}$ junctions becomes enormously stronger than that for the $d_0/d_0$ and $d_0/d_{\pi/4}$ cases.

From Eq.~(4), the asymptotic form of the dissipation kernel is given by
\begin{eqnarray}
\alpha(\tau)
\approx 
\left\{
\begin{array}{cl}
\displaystyle{
  \frac{3^2 \hbar^2}{2^7 \sqrt{2}}
\frac{ R_Q }{\Delta_0 R_N}
\frac{1}{|\tau|^3}
   }
& 
\quad    \mbox{for} \quad \mbox{$d_{0}/d_{0}$}\\
\displaystyle{
  \frac{3\hbar }{2^4 \sqrt{2}}
\frac{ R_Q }{ R_N}
\frac{1}{|\tau|^2}
   }
& 
\quad   \mbox{for}\quad   \mbox{$d_{0}/d_{\pi/4}$}\\
\displaystyle{
  \frac{2^5 \hbar }{35 \pi}
\left(\frac{\Delta_0}{\epsilon}\right)^2
\frac{ R_Q }{ R_N}
\frac{1}{|\tau|^2}
   }& 
\quad   \mbox{for}\quad   \mbox{$d_{\pi/4}/d_{\pi/4}$}
\end{array}
\right.
.
\end{eqnarray}
The result for $d_0/d_0$ junction is in agreement with previous works~\cite{rf:d-waveaction1,rf:d-waveaction2,rf:Bruder,rf:Barash}.

\section{MQT in In-plane $d$-wave Junctions}

Let us move to the calculation of the MQT rate $\Gamma$ for the $d$-wave Josephson junctions based on the standard Caldeira and Leggett theory~\cite{rf:Leggett1}.
At zero temperature $\Gamma$ is given by  
\begin{eqnarray}
\Gamma
\approx
A \exp \left(  - \frac{S_B}{\hbar} \right),
\end{eqnarray}
where $S_B \equiv  S_{eff}[\phi_B]$ and $\phi_B$ is the bounce solution.
Following the procedures as above, we obtain the analytical formulae of the MQT rate for the in-plane  $d$-wave junctions as
\begin{eqnarray}
\! \! 
\frac{\Gamma}{\Gamma_0}
\! \! \approx \! \! 
\left\{
\begin{array}{cl}
     \exp 
    \left[
   -     \left(
     c_0 \frac{3^5 \pi }{2^7 \sqrt{2}}  
       \frac{\hbar \eta }{\Delta_0}
       + 
       \frac{18}{5 }  \frac{ \delta M} {\hbar} 
        \right)
      \frac{U_0}{M \omega_p}
    \right]
& 
  \mbox{for}  \   d_{0}/d_{0}\\
     \exp 
    \left[
   -      \frac{3^4  \zeta (3)}{2^5 \sqrt{2} \pi^2 } \eta  (1- x^2)
    \right]
    &
     \mbox{for}   \    d_{0}/d_{\pi/4}\\
     \exp 
    \left[
   -      \frac{2^8 3^3  \zeta (3)}{35 \pi^3} \left(\frac{\Delta_0}{\epsilon}\right)^2\eta  (1- x^2)
    \right]
    &
       \mbox{for}   \    d_{\pi/4}/d_{\pi/4}
\end{array}
\right.
,
     \nonumber\\
   \! \! \! \! \! 
\end{eqnarray}
where $c_0=\int_0^\infty d y \  y^4 \ln (1 + 1/y^2)/\sinh^2 (\pi y) \approx 0.0135$, $\zeta(3)$ is the Riemann zeta function, $\eta=R_Q/R_N$ is the dissipation parameter,
$U_0$ is the barrier height of the potential $U$, $\omega_p$ is the Josephson plasma frequency, $x=I_{ext}/I_{i}$ (i=1,2,3), and
\begin{eqnarray}
\Gamma_0=12\omega_p 
\sqrt{\frac{
3U_0}{2 \pi  \hbar \omega_p}
  }
\exp 
\left( 
  -
\frac{36 U_0}{5  \hbar \omega_p}
 \right)
\end{eqnarray}
is the MQT rate without the dissipation.
In Eq.~(17) 
\begin{eqnarray}
\delta M 
&=&
\frac{3}{2^4 \sqrt{2}} \frac{\hbar^2 \eta}{\Delta_0}
\int_{-1}^1 d y  \ y^2 \frac{1+y}{\sqrt{1-y}}
\int_0^{\frac{\Delta_0}{\hbar \omega_p}} d z \ z^2 K_1 \left( z |y| \right)^2
.
\nonumber\\
\end{eqnarray}
is the renormalized mass due to the high frequency components ($\omega \geq \omega_p$) of the quasiparticle dissipation.

In order to compare the influence of the ZES and the nodal-quasiparticle on the MQT more clearly, we will estimate the MQT rate (17) numerically.
For a mesoscopic bicrystal YBCO Josephson junction~\cite{rf:mesoYBCO} ($\Delta_0=17.8 \ $meV, $C=20 \times 10^{-15}\ $F, $R_N = 100  \ \Omega$, $x =0.95$), the MQT rate is estimated as 
\begin{eqnarray}
\frac{\Gamma}{\Gamma_0}
\approx
\left\{
\begin{array}{rl}
83 \% & \quad \mbox{for} \quad \mbox{$d_{0}/d_0$} \\
25 \% & \quad \mbox{for} \quad \mbox{$d_{0}/d_{\pi/4}$}\\
0 \% & \quad \mbox{for} \quad \mbox{$d_{\pi/4}/d_{\pi/4}$}
\end{array}
\right.
.
\end{eqnarray}
As expected, the node-to-ZES and ZES-to-ZES quasiparticle tunneling in the $d_{0}/d_{\pi/4}$ and $d_{\pi/4}/d_{\pi/4}$  junctions give strong suppression of the MQT rate compared with the $d_{0}/d_{0}$ junction cases.
Moreover in the $d_{\pi/4}/d_{\pi/4}$ cases, the MQT is almost completely depressed.

\section{Summary}

In conclusion, MQT in in-plane high-$T_c$ superconductors has been theoretically investigate and analytically obtained the formulae of the MQT rate which can be used to analyze experiments.
The node-to-node quasiparticle tunneling in the $d_{0}/d_{0}$ junctions
gives rise to the weak super-Ohmic dissipation as in the case of the $c$-axis junctions~\cite{rf:Kawabata1}.
 For the $d_{0}/d_{\pi/4}$ junctions, on the other hand, we have found that the node-to-ZES quasiparticle tunneling leads to the Ohmic dissipation.
Moreover, in the case of the $d_{\pi/4}/d_{\pi/4}$ junctions, the ZES-to-ZES quasiparticle tunneling gives very strong Ohmic dissipation so the MQT is drastically 
suppressed.

In this paper we have considered the high barrier limit case ($z_0 \gg 1$) only.
In the low barrier cases, the ZES becomes split into two finite energy Andreev levels due to the ZES resonance~\cite{rf:KashiwayaTanaka}.
Moreover, the energy of the split Andreev levels depends on the phase difference $\phi$ and the influence of the proximity effect becomes more important.
To take into account such effects, the present approach should be considerably
modified. 
This issue will be investigated in future articles.
\\

\end{document}